\documentclass{aa}  
\usepackage{multirow}
\usepackage[dvipsnames]{xcolor}
\usepackage{graphicx}
\usepackage{natbib}

\usepackage{txfonts}
\usepackage{placeins}

\usepackage{hyperref}
\hypersetup{
  colorlinks,
  filecolor=blue,
  citecolor=blue,
  linkcolor=blue,
 urlcolor=blue}

\begin{document}

   \title{Gas-Phase metallicity for the Seyfert galaxy NGC~7130}

\author{A.~Amiri\inst{1} 
  \inst{,} \inst{2}   
  \thanks{\emph{ Email:}
     amirnezamamiri@gmail.com}
 \and J.~H.~Knapen \inst{2} \inst{,} \inst{3}
  \and S.~Comerón \inst{3} \inst{,} \inst{2}
  \and A.~Marconi \inst{4} \inst{,} \inst{5}
  \and B.~D.~Lehmer \inst{1}
}

\institute{Department of Physics, University of Arkansas, 226 Physics Building, 825 West Dickson Street, Fayetteville, AR 72701, USA
 \and Instituto de Astrofísica de Canarias E-38205, La Laguna, Tenerife, Spain
  \and Departamento de Astrofísica, Universidad de La Laguna, E-38200, La Laguna, Tenerife, Spain
  \and Dipartimento di Fisica e Astronomia, Universitá degli Studi di Firenze, Via G. Sansone 1, 50019 Sesto Fiorentino, Firenze, Italy
  \and INAF – Osservatorio Astrofisico di Arcetri, Largo E.~Fermi 5, 50125 Firenze, Italy}

\abstract{Metallicity measurements in galaxies can give valuable clues about galaxy evolution. One of the mechanisms postulated for metallicity redistribution in galaxies is gas flows induced by Active Galactic Nuclei (AGN), but the details of this process remain elusive. We report the discovery of a positive radial gradient in the gas-phase metallicity of the narrow line region of the Seyfert~2 galaxy NGC~7130, which is not found when considering the star-forming (SF) components in the galaxy disk. To determine gas-phase metallicities for each kinematic component, we use both AGN and SF strong-line abundance relations, as well as Baldwin–Phillips–Terlevich (BPT) diagnostic diagrams. These relations involve sensitive strong emission lines, namely [O\,{\sc iii}]$\lambda$5007, [N\,{\sc ii}]$\lambda$6584, H$\alpha$, H$\beta$, [S\,{\sc ii}]$\lambda$6716, and [S\,{\sc ii}]$\lambda$6731, observed with the adaptive-optics-assisted mode of the Multi Unit Spectroscopic Explorer at the Very Large Telescope. The presence of a positive radial metallicity gradient only in the AGN ionized component suggests that metals may be transported from central areas to its purlieus by AGN activity.}


   \maketitle
%
\section{Introduction}

Metallicity is one of the most revealing physical quantities in the study of galaxy evolution \citep{bible_of_metallicity}. Metals are produced in galaxies and returned to the interstellar medium (ISM) through a variety of mechanisms, such as supernova explosions \citep{Hillebrandt_supernova,Woosley_supernova}, neutron star mergers \citep{Thielemann_nutron_merging}, and the ejection of gas by asymptotic giant branch stars \citep[e.g.][]{Winckel_AGB}. The gas-phase metallicity of galaxies is impacted by various processes in galaxy evolution, including star formation, gas accretion, gas flows, and wind-driven outflows of gas from galaxies \citep[by supernovae or active galactic nuclei (AGN) activities; e.g.][]{Moustakas_2011_metallicity,cresci_z,Sommariva_2012_stellar_to_gas_z}, and it correlates with physical properties such as the star formation rate \citep{Mannucci_fundamental_Z,lara_fundamental_field_Z}, stellar mass, and morphology \citep{Edmunds_1984}.

In studies of galaxy formation and evolution, the distribution of metals plays an essential role \citep{Sanchez_2014}. Metallicity gradients in galaxies are most often observed to be negative (i.e., decreasing metallicity with increasing radius), but sometimes exhibit positive or flat behaviour. The presence of a negative gradient with a higher metallicity towards the nucleus indicates that star formation begins in the centre of a galaxy and expands outwards \citep{Matteucci}. If galaxies evolve as a closed system and originate from the inside out, negative abundance gradients are expected \citep[e.g.][]{diaz_2005,Franchetto_2020}. This is common in the disks of most spiral galaxies \citep{Kennicutt_2003}. 
Three major surveys have significantly increased the sample of metallicity gradient measurements in the local Universe, namely CALIFA \citep[Calar Alto Legacy Integral Field Area;][]{Sanchez_2012}, MaNGA \citep[Mapping nearby Galaxies at Apache Point Observatory;][]{Bundy_2015}, and SAMI \citep[Sydney-AAO Multi-Object Integral-field spectrograph;][]{Bryant_2015}. The results of these observations reveal that metallicity gradients are primarily negative in nearby galaxies.

A positive metallicity gradient, by contrast, has been observed in several other galaxies \cite[e.g.][]{Cresci_2010,Queyrel_2012,Belfiore_2017,Carton_2018,Wang_2019,Mingozzi_2020,Simons_2021}. 
In a simple scenario, galaxies develop and grow in a dense environment, with gas flowing through, around, and within them \citep{Somerville}. Each of the events that make up this cycle, e.g.~modified star formation, accretion, and mergers, has a unique impact on the galaxy. The central metallicity of galaxies should be diluted, and positive gradients should be observed if metal-poor gas accretion is deposited directly into their centres, resulting in a break in the gradients at small galactocentric distances \citep[e.g.][]{Simons_2021}. \cite{Storchi_metal_poor_bh_feed} show that the capture of a gas-rich dwarf galaxy, which is a process that can start nuclear activity in galaxies, can result in the accretion of metal-poor gas into the nuclear region and its dilution. In a high-redshift investigation of star-forming (SF) galaxies, \cite{Queyrel_2012} observed positive metallicity gradients and proposed that the inversion of the abundance gradient might be caused by interactions with the environment.

Metallicity gradients can also be flat. For example \cite{Nascimento_manga} demonstrated that almost
10\% of their AGN from MANGA are consistent with a constant metallicity across the galactic disc. 

In addition to star formation, stellar evolution, and environmental influences, the presence of an AGN may have an impact on the evolution of the host galaxy properties \cite[e.g.][]{grove_2006,Coil_2015,Thomas_2019,Armah_2023}. 
The metal enrichment due to AGN could be due to either an in situ top-heavy initial mass function (IMF) in the accretion disk around the supermassive black hole \cite[e.g.][]{Nayakshin_2005} or dust destruction in the broad line region (BLR), which releases metals into the ISM \citep{bible_of_metallicity}. In this case, following \cite{bible_of_metallicity}, the AGN would promote fast star formation and ISM enrichment.

Ionized gas outflows detected through optical emission lines are frequently observed in the narrow line regions (NLRs) and the extended NLRs \cite[e.g.][]{Veilleux, unger_1987, Pogge1988}. NLR outflows present valuable information for studying the interaction between AGN and their host galaxies. The high resolution capabilities of modern telescopes can resolve the NLRs of nearby galaxies within just a few parsecs of the supermassive black hole (SMBH), and observe their extension over several kiloparsecs into the bulges and/or disks of the host galaxies \citep{Kang_2018}. The kinematics and physical conditions of the ionized gas in the NLR provide unique information regarding the properties of the outflows, including the energy associated with them \cite[e.g.][]{Bennert_2006,Vaona, Dopita_2014_nlr,Chen_2019,Zhang_2022,Meena_2023}. The NLR would then be further enhanced by AGN-driven outflows of high-metallicity gas that has been shown to be ejected from the BLR on kiloparsec (kpc) scales \cite[e.g.][]{DOdorico}. In-situ star formation within AGN-driven outflows is another potential contribution to metal enrichment of the gas around the BLR \cite[e.g.][]{Maiolino_2017,Gallagher_2019}.

In this paper we study, for the first time, the metallicity gradients separately in the disc, low velocity dispersion, and high velocity dispersion components of NGC~7130 by using the multi-component fits of the emission lines performed in \citet{sebastien_ngc7130}. The paper is organized as follows. In Section~2, we provide a brief overview of the galaxy areas used in our study. In Section~3, we outline how we classify the data into AGN and SF regions. We also elaborate on the calibration relations employed to estimate the gas-phase metallicities and their variations as a function of radial distance. Finally in Section~4, we discuss our findings in the conclusion, summarizing the implications and significance of our study.
\section{Observational data}

\label{sect2}

The southern galaxy NGC~7130 has a redshift of $z=0.016151$. With an axial ratio of 0.88, this galaxy is almost face-on \citep{Skrutskie_2006}. It is a peculiar Sa galaxy \citep{devacouleurs_1991} with a bar \citep[e.g][]{Mulchaey_1997,Malkan_1998,Dopita_2002,Mar_2010,Zhao_2016}, and shows evidence of an ionised outflow \citep{Knapen_2019}. NGC~7130 also hosts a Seyfert~2 AGN \citep{Phillips}. Radio and optical investigations suggest that both star formation and nuclear activity contribute to ionising the gas in the nuclear region.

The optical spectrum of NGC~7130 shows narrow (low velocity dispersion, $\sigma<250\,{\rm km\,s^{-1}}$) and broad kinematic components (high velocity dispersion, $\sigma>250\,{\rm km\,s^{-1}}$) that correspond to both ambient and outflowing gas in the galaxy \citep{Knapen_2019,sebastien_ngc7130}. These authors use optical integral field spectroscopy at high angular resolution ($7.5\times7.5$ arcsec$^2$ field of view and the point source function has a full width at half maximum of about 0.18 arcsec) obtained with the adaptive-optics-supported the Multi Unit Spectroscopic Explorer (MUSE) instrument on the European Southern Observatory (ESO) Very Large Telescope (VLT).

In this study, we use information from the multi-component decomposition of the circumnuclear ISM of NGC~7130 from \citet{sebastien_ngc7130}, which was based on the principal MUSE-wavelength emission lines. Each of the spectral lines was fitted with a superposition of up to six Gaussian components with distinct velocities and velocity dispersions. The fit was made using the \texttt{python} re-implementation of the \texttt{GandALF} software \citep{sarzi2006,Falcon2006} \textbf{(}called \texttt{pyGandALF}\textbf{)} \citep{Bittner_2019} over 2689 spectral bins with a signal-to-noise ratio of the H$\alpha$ line of 100 or more generated using the Voronoi binning algorithm by \citet{Cappellari2003}. To reduce the number of free parameters, the kinematics of different emission line components were tied to those of the H$\alpha$ line. The number of components required for a given spectrum was chosen based on criteria using the $\chi^2$ goodness-of-fit estimator \citep[see flow chart in Fig.~3 from][]{sebastien_ngc7130}. In total, nine distinct kinematic components were characterised. 

Six of the nine kinematic components found by \citet{sebastien_ngc7130} are associated with the AGN outflow. Their nature is deduced from their large velocity with respect to the galaxy (typically 100s of ${\rm km\,s^{-1}}$), their relatively large velocity dispersion (of 100s of ${\rm km\,s^{-1}}$ or more), and line ratios incompatible with an ionization by stars. These components related to the AGN can be further subdivided into kinematically narrow components (typical velocity dispersion below $250\,{\rm km\,s^{-1}}$) and kinematically broad components (typical velocity dispersion above $250\,{\rm km\,s^{-1}}$). A further two of the nine kinematic components of the circumnuclear medium in NGC~7130 are related to the disc. The final component might be an artifact caused by the fitting procedure.

Based on their observations, \citet{sebastien_ngc7130} produce a toy model of the circumnuclear medium of NGC~7130 (see their Fig{\bf .}~14). They propose that the ISM in the disc has a low-density ($n_e\approx90\,{\rm cm}^{-3}$), low velocity dispersion ($\sigma<100\,{\rm km\,s}$) background on top of which we observe a larger density ($n_e\approx500\,{\rm cm}^{-3}$), slightly blueshifted ($-100\,{\rm km\,s^{-1}}<V<0\,{\rm km\,s^{-1}}$), and slightly higher velocity dispersion ($\sigma<250\,{\rm km\,s^{-1}}$) gas associated with the strongest knots of star formation  and that was called the zero-velocity narrow component to distinguish it from kinematically narrow components in the outflow. The AGN outflow is postulated to have a biconical morphology with a main axis that is close to coplanar with the galaxy disc. Each of the cones would be made of one or more collimated narrow  components, surrounded by a broad component with a larger opening angle. The blueshifted components would correspond to the approaching side of the bicone (located slightly above the plane of the galaxy), whereas the redshifted ones would correspond to the receding one (located below the mid-plane). According to the model, the redshifted side of the cone is partly obscured by extinction from the disc, and hence parts of it are not observed or are seen at a lower signal-to-noise level than their blueshifted counterparts. Hence, the properties derived for the redshifted components of the outflow are more uncertain than those derived for the blueshifted components. 

In this study, we consider four components out of nine from \citet{sebastien_ngc7130}, that is the gas-phase metallicity in the disc, blueshifted broad, blueshifted narrow, and zero velocity narrow components. The reason to exclude the redshifted high velocity dispersion and the redshifted low velocity dispersion components is ultimately that the redshifted components are obscured by the disc, so it is harder to obtain precise gas-phase metallicity measurements. The zero velocity high velocity dispersion component is probably a spurious artifact required to fit the spectra in some of the bins. We also do not investigate the crescent low velocity dispersion component because this region combines the interaction of the jet and disk and there is no clear method to calculate metallicity in those complex regions.

\section{Results}

\begin{figure*}
    \includegraphics[width=\linewidth]{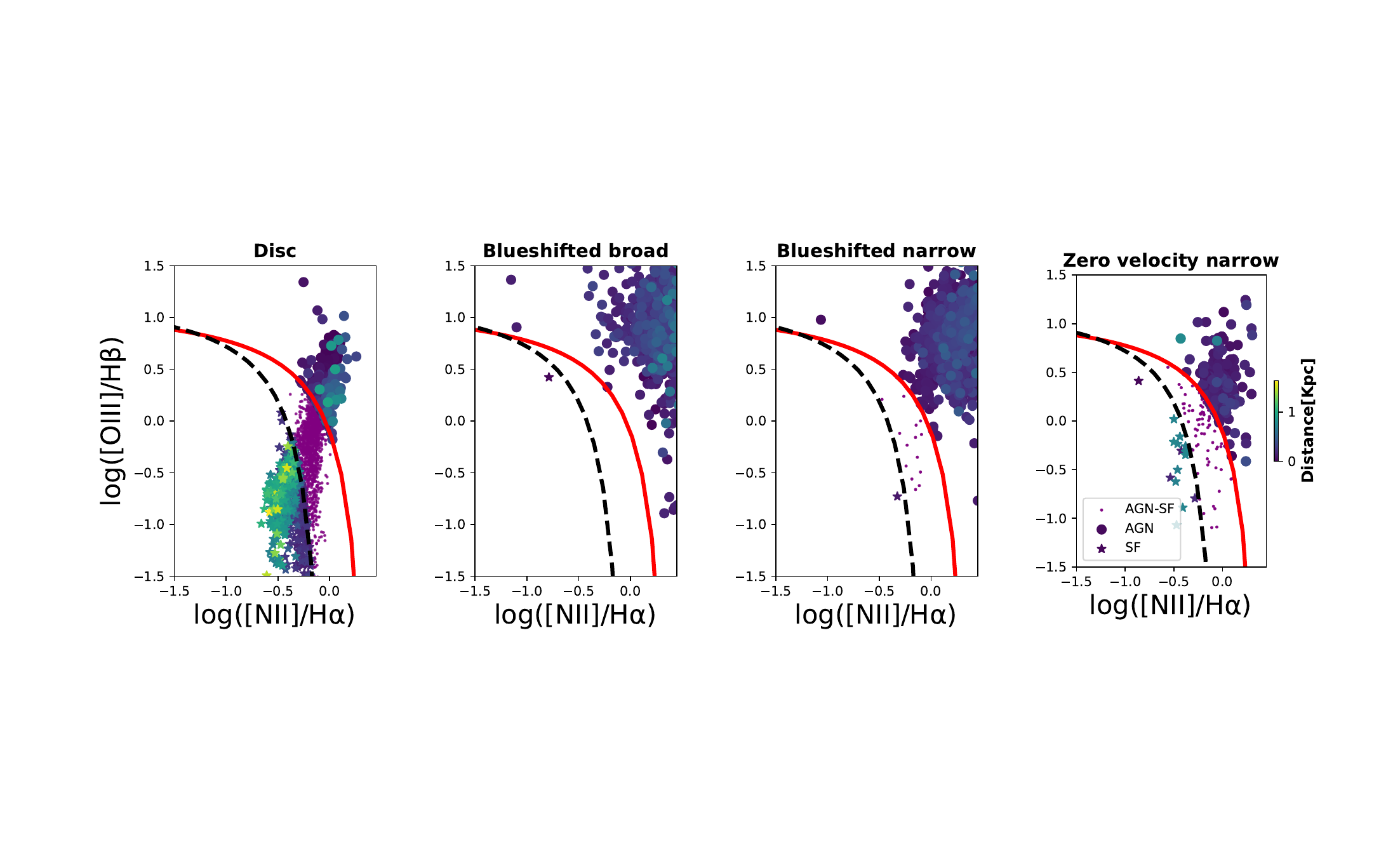}
    \caption{BPT diagnostic diagram to discriminate between SF ionised and AGN-ionised regions. The dashed black line and the solid red line show the separations between the two populations according to the boundaries from \citet{Kewley_bpt} and \citet{kauffmann_bpt}, respectively. Circles and blue stars refer to AGN and SF dominated regions, respectively. Plots are colour-coded according to how far bins are from the centre. We have not investigated the regions located between the two lines (purple circles).}
              
    \label{fig1}
\end{figure*}

To classify regions in NGC~7130 as AGN-ionized or H\,{\sc ii}-ionized, we use the standard Baldwin–Phillips–Terlevich (BPT) diagrams \citet{bpt_main} applied to the spectra of each of the Voronoi bins defined by \citet{sebastien_ngc7130}. The distribution of the {bins} in the [O\,{\sc iii}] 5007/H$\beta$ versus the [N\,{\sc ii}]6584/H$\alpha$ diagram, together with the boundary lines between SF and AGN defined by \citet{Kewley_bpt} and \citet{kauffmann_bpt} are shown in Fig.~\ref{fig1}. The red demarcation line by \citet{Kewley_bpt} is a theoretical upper limit on the location of SF galaxies in this diagram, obtained using a combination of photoionization and stellar population synthesis models. It yields a conservative selection of AGN. \citet{kauffmann_bpt} revised the boundary line on the basis of observational evidence that SFs and AGNs are distributed along two separate sequences. It yields a conservative SF selection. In order to avoid ambiguous classifications we have adopted the more conservative selection for both SFs and AGNs, excluding from further consideration the \textbf{bins} located between the two lines. These could include regions with a mixture of ionizing sources \citep[e.g.][]{Allen,rosa_interact_z}, and are expected to have a mixed AGN-stellar emission as their ionizing source \citep[e.g.][]{Davis_2014}. In Fig.~\ref{fig1}, the diagnostic BPT diagram displays the distribution of AGN and SF bins through each component. We find that the disc component is partly ionized by SF and AGN regions, whereas the other three components (outflow) are dominated by AGN photoionization.

\subsection{Gas-phase metallicity estimation}
The gas-phase metallicity in each bin is calculated using calibrations based on strong emission lines, i.e. adopting the so-called strong-line method. The MUSE wavelength coverage and the laser-affected wavelength (contamination caused by laser during adaptive optics observations) prevent us from measuring the temperature-sensitive emission lines, [O\,{\sc  iii}]$\lambda$4363 and [N\,{\sc ii}]$\lambda$5756, and thus from estimating element abundances using the direct method ($T_{e}$). While we are able to identify [O\,{\sc ii}]$\lambda$$\lambda$7319,7330 in the majority of bins, the doublet [O\,{\sc ii}]$\lambda$$\lambda$3720,3730 fell outside the MUSE wavelength range, making it impossible to calculate abundances based on these auroral lines. We are aware of the systematic discrepancies in gas-phase metallicities which are potentially present when using strong-line methods: differences up to 0.6 dex for H\,{\sc ii} regions \citep[e.g.][]{ Kewley_2008_discrepancy,Sanch_discrepancy,Guerrero_2012_discrepancy} and up to 0.8 dex for AGN can be found when comparing metallicities obtained using the strong line method based on calibrations from different authors,  particularly in the low-metallicity regime (smaller than 8.5, e.g., \citealt{Dors_2020_b_discrep}).

To estimate gas-phase metallicities we utilize the emission-line intensities [O\,{\sc iii}]$\lambda$5007, [N\,{\sc ii}]$\lambda$6584, H$\alpha$, H$\beta$, [S\,{\sc ii}]$\lambda$6716 and [S\,{\sc ii}]$\lambda$6716. We discard the bins where one or more of the aforementioned emission lines are not detected. We consider two different calibration relations to estimate the gas-phase metallicities for SF- and AGN-dominated regions. Hereafter, we refer to gas-phase metallicity as $Z_{\rm gas}$.
For SF regions, we utilized the methodology developed by \citet{curti_stack} to determine the gas-phase metallicity. \citet{curti_stack} introduced novel empirical calibrations specifically designed for a selection of commonly employed strong-line diagnostics and the scatter around the calibration varies up to 0.15. These calibrations enable accurate calculation of the oxygen abundance in star forming galaxies and allow us to estimate the gas-phase metallicity in SF regions. Following \citet{curti_stack}, we define:
\begin{eqnarray} \label{mirko_3}
{\rm N2} & = &  {[{\rm N}\,\textsc{ii}]\lambda6584/{\rm H}\beta},\\
\nonumber\\
\label{mirko_4}
{\rm S2} &=& ({[{\rm S}\textsc{ii}]\lambda6716+[{\rm S}\textsc{ii}]\lambda6730)/{\rm H}\beta},\\
\nonumber\\
\label{mirko_5}
{\rm R3} &=& {[{\rm O}\textsc{iii}]\lambda5007/{\rm H}\beta},
\end{eqnarray}
and calculate the gas-phase metallicity for a given SF region following:

\begin{equation} \label{mirko_1}
\begin{split}
Z_{\rm gas} = 8.424+ (0.030\times \log_{10}(1.33\times {\rm R3}/{\rm S2})) \\
+0.751\times \log_{10}(1.33\times {\rm N2})+\\
(-0.349+0.182\times \log_{10}(1.33\times {\rm R3}/{\rm S2})+ \\
0.508\times \log_{10}(1.33\times {\rm N2}))\times \log_{10}({\rm S2})
\end{split}
\end{equation}
for $\log_{10}({\rm N2}) > -0.6$ and \\
\begin{equation} \label{mirko_2}
\begin{split}
Z_{\rm gas} = 8.072+ 0.789\times \log_{10}({\rm R3}/{\rm S2})\\
+0.726\times \log_{10}({\rm N2})+ (1.069 - 0.170\times \log_{10}({\rm R3}/{\rm S2}) + \\
0.022\times \log_{10}({\rm N2}))\times \log_{10}({\rm S2})
\end{split}
\end{equation}
for $\log_{10}({\rm N2}) \leq -0.6$. 
\\

To compute gas-phase metallicities in AGN regions, we use the relation by \citet{Storchi-Bergmann_1998}. They proposed first a calibration between the metallicity $Z_{\rm gas}$ and the intensities of optical emission-line ratios of AGNs that is valid for the gas-phase metallicity in the range $8.4\leq Z_{\rm gas}\leq 9.4$. The computed $Z_{\rm gas}$ from these calibrations varies by $\sim0.1\,{\rm dex}$. The metallicity value should be corrected in order to take into account electron density ($N_{\rm e}$) effects:
\begin{equation} \label{berg1}
Z_{\rm gas} = {Z_{\rm int} - (0.1\times \log_{10}(n_{e} /300\,{\rm cm}^{-3})}
\end{equation}
in which:
\begin{equation} \label{berg2}
\begin{split}
Z_{\rm int} = 8.34 +(0.212\times {\rm N2})
-(0.012 \times {\rm N2}^{2})-(0.002\times {\rm R3})
\\+ (0.007\times ({\rm N2}\times R3))- (0.002\times {\rm N2}^{2}\times {\rm R3})\\ + (6.52\times (10^{-4}\times {\rm R3}^{2}))
+ (2.27\times 10^{-4}\times ({\rm N2}\times {\rm R3}^{2})) \\+ (8.87\times 10^{-5} \times ({\rm N2}^{2} \times {\rm R3}^{2})).
\end{split}
\end{equation}

\begin{figure}
    \includegraphics[width=9cm]{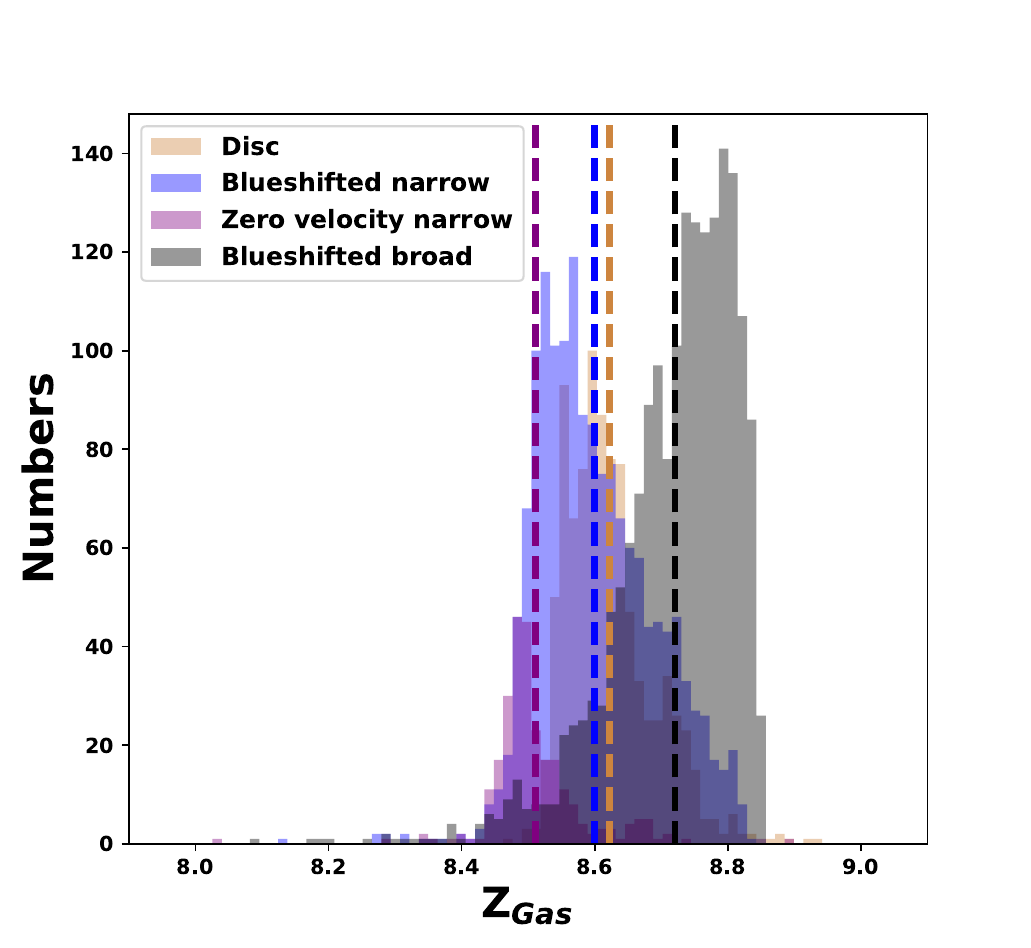}
    \caption{Distribution of $Z_{\rm gas}$ for the four components described Sect.~\ref{sect2}. The vertical dashed lines show the median value of $Z_{\rm gas}$ for each of the components.}
              
    \label{fig2}
\end{figure}

\begin{figure*}
    \centering
    
    \vspace{1mm}
    
    \includegraphics[width=\linewidth]{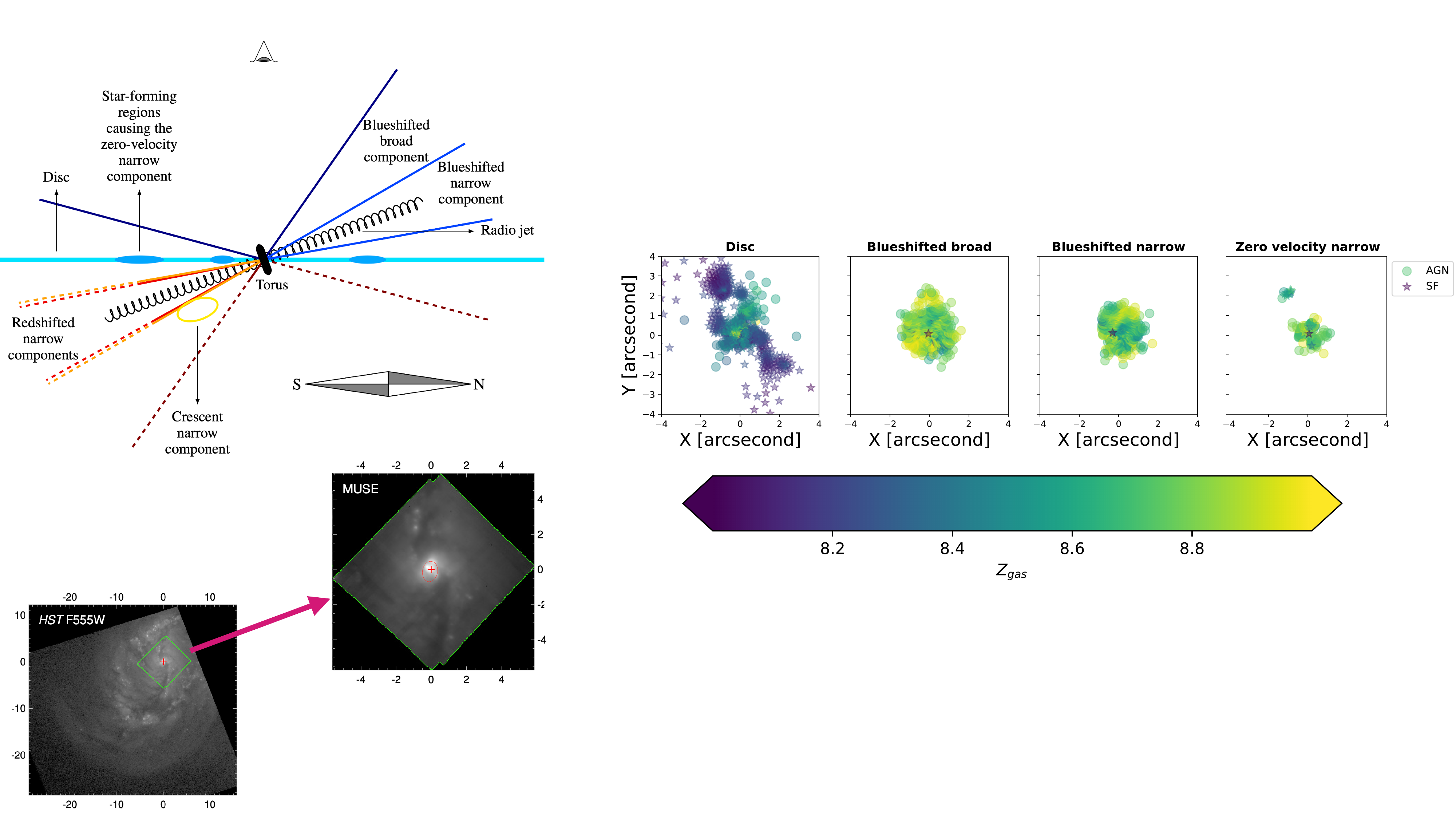}
    \caption{Hubble Space Telescope F555W image of the central region of NGC~7130 superimposed with the MUSE-NFM coverage area and the MUSE data cube integrated along the spectral direction (lower-left panel). The top-left panel demonstrates and illustrates the ionised circumnuclear gas in NGC~7130 using the toy model from \citet{sebastien_ngc7130}. The panel-right displays the spatial (X,Y) distribution of bins for four of the nine components in NGC~7130. Based the BPT classification, the circles demonstrate AGNs while stars show SFs. The coordinate labels on each panel are presented in arcseconds and the colour bar shows $Z_{\rm gas}$.}
              
    \label{fig3}
\end{figure*}

To estimate the $N_{\rm e}$, we adopt the \citet{sebastien_ngc7130} measurements, which mainly used the total [S\,{\sc ii}]]$\lambda$6716 and [S\,{\sc ii}]$\lambda$6730 flux ratio calibration from \citet{Sanders_2016}. Each component's variations in $Z_{\rm gas}$, the combination of both AGNs and SFs, result in the histogram shown in Fig.~\ref{fig2} while Fig.~\ref{fig3} shows the spatial distribution (X,Y) of AGN and SF bins. We exclude a small number of data points (less than ten bins) for each component, where $Z_{\rm gas}\leq8.2$.

\subsection{Radial variations of $Z_{\rm gas}$}

There is an increasing amount of observational data supporting the existence metal-rich outflows from AGN-powered winds and/or jets  \citep{McNamara_2012}. Cold gas that is highly enriched in metals has been discovered recently in various clusters and groups \citep[e.g.][]{Mishra2022}. This gas is either correlated with radio jets or found preferentially along cavities. Its abundance is generally near-solar or even super-solar, and in some cases, it is even more enriched than the regions that are closest to the centre of galaxies \cite[e.g.][]{Simionescu,OSullivan_2011,Kirkpatrick_2}. Given the large amount of energy needed to move gas to its measured position, it is assumed that this metal-enriched gas was carried to its observed location by AGN-related processes \cite[e.g.][]{Simionescu_ff,Pope,Kirkpatrick_ff}.

\begin{figure*}
    \includegraphics[width=\linewidth]{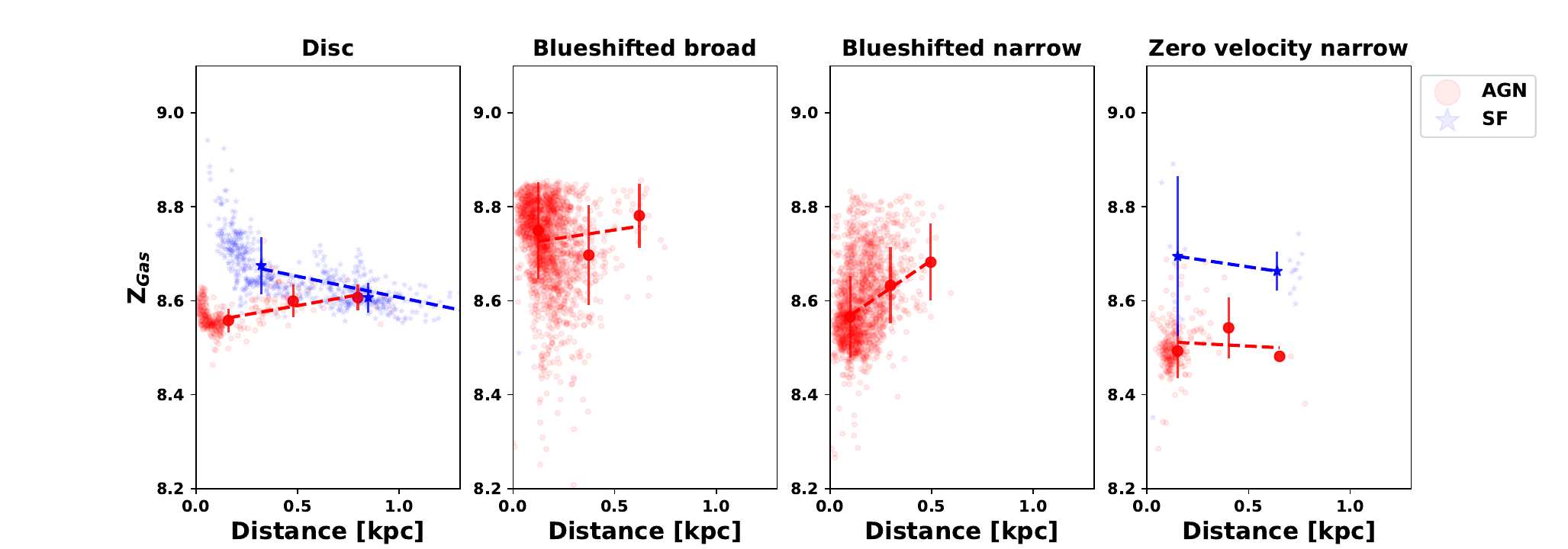}
    \caption{Metallicity gradient profiles for each component using the \citet{Storchi-Bergmann_1998} and \citet{curti_stack} metallicity relations for AGNs (red circles) and SFs (blue stars), respectively, with radius in kpc on the bottom axis. Over plotted are binned linear fits to the radial metallicity gradient in AGNs (dotted red lines) and SFs (dotted blue lines). The larger markers demonstrate binned median  values, and error bars represent the scatter in the data within each bin.}
              
    \label{fig4}
\end{figure*}

We derive radial chemical abundance profiles for the different components of NGC~7130, as shown in Fig.~\ref{fig4}. We employed a binned linear regression approach to fit the radially averaged $Z_{\rm gas}$ as a function of radial distances. In Table~\ref{tab:caption}, we list the slopes and intercepts of the gradients down to the centre ($R=0$) for each component. For the blueshifted narrow and blueshifted broad components, we find an overall positive radial trend which suggests a gradual increase in gas-phase metallicity as we move towards the outer regions. This finding implies that there must exist strong internal mechanisms within the galaxy that actively transport high amounts of metals towards these outer areas.

\begin{table*}
\caption{The radial distribution results of $Z_{\rm gas}$}
\begin{tabular}{ccccccccccc}

    \hline
    \multirow{2}{*}{Components} & \multicolumn{5}{c}{AGNs} & \multicolumn{5}{c}{SFs} \\
    & Slope & intercept & $r$-value & $p$-value & stderr& Slope & intercept & $r$-value & $p$-value & stderr\\
    \hline
    Disc & $0.07$ & $8.55$ & $0.92$ & $0.24$& 0.03& $-0.08$ & $8.69$ & $-0.96$ &$0.15$ &$0.02$   \\
    Blueshifted broad  & $0.06$ & $8.71$ & $0.37$ & $0.75$& $0.15$& $-$ & $-$ & $-$ &$-$ &$-$ \\
    Blueshifted narrow  & $0.29$ & $8.53$ & $0.99$ & $0.05$& $0.02$& $-$ & $-$ & $-$ &$-$ &$-$ \\
    Zero-velocity narrow & $-0.02$ & $8.51$ & $-0.17$ & $0.88$& $0.12$& $-0.06$ & $8.70$ & $-1$ &$0$ &$0$ \\
    \hline

\end{tabular}
\label{tab:caption}
\tablefoot{General form of equation to describe $Z_{\rm gas}$ variations as a function of radius is: $Z_{\rm gas}=({\rm Slope}\times{\rm distance[kpc]})+{\rm intercept}$.}
\end{table*}

The disc component (top-left panel in Fig.~\ref{fig3} and left panel in Fig.~\ref{fig4}) is harder to interpret, as it shows a superposition of AGN- and SF-ionized regions. Since the axis of the AGN outflow is probably close to the plane of the disc \citep{sebastien_ngc7130}, it is possible that the AGN radiation is genuinely ionizing parts of the disc. However, it is also possible that the bins ionized by the AGN are an artefact caused by the difficulty of measuring the intensity of the component of the [O\,{\sc iii}] lines corresponding to the disc \citep[see Fig.~A1 and A2 from][to appreciate how these can be easily buried under the components corresponding to the outflow]{sebastien_ngc7130}. The parts of the disc ionized by the AGN tend to show a positive gradient of the radial gas-phase metallicity.

For the SF bins in the disc (the zero velocity narrow component), we also find a negative radial gas-phase metallicity. Our results align with the notion that galaxies underwent relatively smooth gas accretion histories, with metal-poor inflows and outflows preferentially affecting the outer regions of galaxies. This combined with the inside-out evolution of galaxies, naturally gives rise to negative metallicity gradient \cite[e.g.][]{Boardman}.

\section{Conclusions}
We address the still unsettled issue of the radial distribution of gas-phase metallicities in AGN, by exploring the different gas components identified in  NGC~7130 by \cite{sebastien_ngc7130}. We distinguish between SF bins and bins dominated AGNs by means of the classical BPT diagnostics. Depending on the type of emission lines that may be detected in both AGN and SF regions, many studies have proposed varying strong line ratios to characterize the metallicities \cite[e.g.][]{Carvalho_agn_calibration,Storchi-Bergmann_1998, perez_montero,curti_stack}. Depending the location in the BPT diagram, we compute gas-phase metallicities based on either the \cite{curti_stack} or the \cite{Storchi-Bergmann_1998} calibration relations.  We then analyse the radial gas-phase metallicity distribution for each component, separately.

The majority of the gas-phase metallicities that we observe in the AGN component are consistently increasing with radius, meaning that the activity of the AGN indeed plays a significant role in shaping the radial distribution of the gas-phase metallicity. This suggests that AGN activity is responsible for actively transporting metals from the central region to its purlieus.
This intriguing phenomenon has the potential to establish a notably steep relationship between the radial distance from the galactic centre and the gas-phase metallicity. 

We find that for the AGN-ionized gas $Z_{\rm gas}$ has the lowest values specifically at the nuclear region. This piece of information potentially holds key insights into the environmental influences. The present AGN activity of NGC~7130 galaxy may be the result of a recent accretion from a metal-poor dwarf galaxy. Although there is no strong observational evidence of interaction between NGC~7130 and its neighbouring galaxies, the warped aspect of the outskirts of NGC~7130 \citep{1976srcb.book.....D} may demonstrate a past close encounter between them. Further signs of a possible recent interaction are the asymmetric velocity and velocity dispersion maps of the ionized gas in the galaxy \citep{Bellocchi_2012}. As \citet{Storchi_metal_poor_bh_feed} have shown, the capture of a gas-rich dwarf galaxy, which is a process that can start nuclear activity in galaxies, can result in the accretion of metal-poor gas into the nuclear region and its dilution.

To conclude, our findings emphasize the crucial role that AGN activity plays in shaping the metal enrichment of galaxies and provide valuable insights into the underlying processes driving the gas-phase metallicity gradients in galaxies. Our work highlights the importance of internal mechanisms in redistributing metal content throughout a galaxy from centre to outskirt.

\begin{acknowledgements}
A.A. thanks Kastytis Zubovas, C. Ramos Almeida, Rogério Riffel, and A.Khoram for helpful discussions. Also, A.A. acknowledges support from the ACIISI, Consejer\'{i}a de Econom\'{i}a, Conocimiento y Empleo del Gobierno de Canarias and the European Regional Development Fund (ERDF) under the grant with reference PROID2021010044.
SC acknowledges funding from the State Research Agency (AEI-MCINN) of the Spanish Ministry of Science and Innovation under the grant $'$Thick discs, relics of the infancy of galaxies$'$ with reference PID2020-113213GA-I00.
Co-funded by the European Union (MSCA EDUCADO, GA 101119830). Views and
opinions expressed are however those of the author(s) only and do not
necessarily reflect those of the European Union. Neither the European
Union nor the granting authority can be held responsible for them. JHK
acknowledges grant PID2022-136505NB-I00 funded by MCIN/AEI/10.13039/501100011033 and EU, ERDF.

\end{acknowledgements}

\bibliographystyle{aa}
\bibliography{Z_NGC7130}

\end{document}